\documentclass[acmsmall,nonacm]{acmart}\usepackage[]{graphicx}\usepackage[]{color}
\makeatletter
\def\maxwidth{ %
  \ifdim\Gin@nat@width>\linewidth
    \linewidth
  \else
    \Gin@nat@width
  \fi
}
\makeatother

\definecolor{fgcolor}{rgb}{0.345, 0.345, 0.345}

\usepackage{framed}
\makeatletter
 {\par\unskip\endMakeFramed%
 \at@end@of@kframe}
\makeatother

\definecolor{shadecolor}{rgb}{.97, .97, .97}
\definecolor{messagecolor}{rgb}{0, 0, 0}
\definecolor{warningcolor}{rgb}{1, 0, 1}
\definecolor{errorcolor}{rgb}{1, 0, 0}
\usepackage{alltt}

\usepackage{subcaption}

\AtBeginDocument{%
  \providecommand\BibTeX{{%
    \normalfont B\kern-0.5em{\scshape i\kern-0.25em b}\kern-0.8em\TeX}}}

\setcopyright{none}
\acmBooktitle{}
\IfFileExists{upquote.sty}{\usepackage{upquote}}{}
\begin{document}

\title{Software-Supported Audits of Decision-Making Systems: Testing Google and Facebook's Political Advertising Policies}

\author{J. Nathan Matias}
\affiliation{
    \institution{Cornell University}
    \city{Ithaca}
    \state{New York}
    \country{United States of America}
}
\email{nathan.matias@cornell.edu}

\author{Austin Hounsel}
\affiliation{
    \institution{Princeton University}
    \city{Princeton}
    \state{New Jersey}
    \country{United States of America}
}
\email{ahounsel@cs.princeton.edu}

\author{Nick Feamster}
\affiliation{
    \institution{University of Chicago}
    \city{Chicago}
    \state{Illinois}
    \country{United States of America}
}
\email{feamster@uchicago.edu}

% \renewcommand{\shortauthors}{J. Nathan Matias, Austin Hounsel, and Nick Feamster}

\begin{abstract}
    How can society understand and hold accountable complex human and algorithmic decision-making systems whose systematic errors are opaque to the public?
    These systems routinely make decisions on individual rights and well-being, and on protecting society and the democratic process. Practical and statistical constraints on external audits--such as dimensional complexity--can lead researchers and regulators to miss important sources of error in these complex decision-making systems.
    In this paper, we design and implement a software-supported approach to audit studies that auto-generates audit materials and coordinates volunteer activity.
    %software for overcoming these constraints by auto-generating audit materials and coordinating volunteer audit activity.
    We implemented this software in the case of political advertising policies enacted by Facebook and Google during the 2018 U.S. election.
    Guided by this software, a team of volunteers posted 477 auto-generated ads and analyzed the companies' actions, finding systematic errors in how companies enforced policies.
    We find that software can overcome some common constraints of audit studies, within limitations related to sample size and volunteer capacity.
\end{abstract}




\maketitle

% The file aaai21.sty is the style file for AAAI Press

% Title

% Your title must be in mixed case, not sentence case.
% That means all verbs (including short verbs like be, is, using,and go),
% nouns, adverbs, adjectives should be capitalized, including both words in hyphenated terms, while
% articles, conjunctions, and prepositions are lower case unless they
% directly follow a colon or long dash

\section{Introduction}
In 2018, both Facebook and Google introduced policies that restricted who could purchase advertisements about elections and topics of ``national importance'' \cite{leathern_shining_2018, wakabayashi_google_2018}. During the 2016 U.S. presidential election, foreign actors had attempted to influence U.S. voters by purchasing ads, among other tactics \cite{isaac_russian_2017}. Under pressure from regulators and civil society, the companies needed a way to review potentially millions of ads a week, identify which ones were related to an election, and determine which should be published---all without disrupting ad markets and other forms of societally-important speech~\cite{cspan2017facebook,hill2017rubles,isaac_russian_2017}. They needed large-scale decision-making systems.

These systems also make mistakes. In the 2018 midterm Congressional elections ("midterms"), Facebook was accused of wrongly blocking ads for community centers, news articles, veterans pages, and food advertising. Enough mistakes could substantially impact the civic fabric of democracies. Public holidays, community centers, and news conversations help communities understand each other and work together effectively \cite{putnam_bowling_2000}. When platforms wrongly restrict a community's capacity to connect, this civic fabric weakens, with less participation in public life, fewer relationships among neighbors, and less common understanding.

More broadly, decision-making systems, which often combine human and software processes, routinely determine high-stakes outcomes for millions of people. These systems shape people's lives in areas including judicial sentencing \cite{angwin_machine_2016}, child protection \cite{chouldechova_case_2018}, public health \cite{ card_how_2018, chen_can_2019},
%insurance pricing \cite{julia_angwin_minority_2017},
immigration \cite{whittaker_u.s._2019}, and hiring \cite{barabas_engineering_2015}. Decision-making systems are used worldwide to protect the public from harmful speech \cite{gillespie_custodians_2018}, to protect democracies from attempts to undermine elections \cite{madrigal_will_2018}, and to censor political speech \cite{king_reverse-engineering_2014}. Because all decision-making systems make errors, the public depends on independent audits to hold decision-making institutions accountable, correct errors, and set policy for their decisions \cite{popper_open_1947, diakopoulos_algorithmic_2014}.

Attempts to audit decision-making systems can also advance scientific understanding of society, algorithms, and how they interact in the field \cite{bandiera_field_2011, kitchin_thinking_2017, rahwan_machine_2019}. Audit studies, unlike financial audits or qualitative audits \cite{maclay_protecting_2010, sandberg_second_2019}, are randomized trials that quantitatively estimate systematic statistical error in decision-making systems. By conducting lab and field experiments that observe systematic errors in decision-making, scientists have advanced understanding of human psychology \cite{kahneman_subjective_1972}, organizational behavior \cite{pager_use_2007}, and market behavior \cite{salganik_web-based_2009}. Computer scientists use similar methods to study decision-making by algorithms \cite{chen_observing_2017, sandvig_auditing_2014}.

Because audit studies are complex to design and coordinate, researchers tend to focus on a small number of decision-making error types. But if they are too simple, studies can fail to reveal important sources of error \cite{adler_auditing_2018}. In this paper, we describe, prototype, and evaluate novel software for increasing the complexity and validity of audit studies. Like other computer science papers that introduce a new approach to software-supported research \citep{geiger_trace_2011, matias_civilservant_2018}, we present the method and include an example of its use. We created software to (a) support variable selection, (b) generate audit materials, (c) allocate audit attempts to testers, and (d) generate pre-registered statistical results and illustrations.

Using this software, we conducted a comparative audit of Google and Facebook's political advertising policies during the 2018 U.S. midterm election. In our study, testers posted 477 auto-generated ads across the two platforms, testing each platform's policy decisions for different ad targeting criteria, for partisan bias across multiple kinds of content, and for different advertiser characteristics.
While Google did not make any mistakes in our audit, Facebook mistakenly prohibited 17.5\% of ads for government websites and 4.9\% of ads to non-election civic events.
In total, Facebook mistakenly prohibited 4.2\% of our ads. We did not find evidence of political bias in these mistakes.

Overall, this paper explains audit studies as a design challenge for computer scientists. By reporting an audit study conducted with working software, we demonstrate the potential for software to streamline the design and implementation of audit studies while expanding their realism and validity. We conclude with design implications for software-supported audits of decision-making systems.

\section{Background and Related Work}
We begin by providing background on audit studies and introduce limitations that software can help overcome. We then provide background on the political advertising policies of Facebook and Google during the 2018 U.S. midterm elections to illustrate these challenges in the field. Finally, to frame our discussion of design considerations for audit software, we review the state of methods for software-supported audits.

\subsection{Audit Studies and Their Limitations}

Audit studies are field experiments that describe and explain systematic error made by a decision-making system. These randomized trials, first developed in the 1970s \cite{wienk_measuring_1979, hakken_discrimination_1979}, became more common in the 2000s after a series of field experiments by Devah Pager demonstrated systematic racial discrimination by U.S. employers \cite{pager_use_2007, pager_mark_2003}. 
%While social scientists and advocates have persuasively argued the shortcomings of a focus on decision-making systems when addressing social inequality \cite{alexander_new_2010}, audit studies remain a valuable method for studying errors by a wider range of decision-making systems.

\subsubsection{How Audit Studies Work}
In audit studies, researchers start with hypotheses about decision-making error based on the (a) characteristics of the "tester" (i.e., a study participant that prompt the decision-making system) and (b) the choice being offered to the decision-making system. To create a controlled test, researchers recruit testers who are as similar as possible on all characteristics except those of interest to the study (e.g., skin tone). Testers are then randomly assigned to prompt a decision-making system to make a choice, for example by applying for a job. Across multiple prompts, testers also vary the stimulus provided to the institution (e.g., altering the CV) and record the decision made by the system. Researchers then conduct statistical tests for differences in decision rates for different categories of testers and prompts.

\subsubsection{Dimensional Complexity}
Audit studies, like all randomized trials, increase in dimensional complexity with each added variable. Pager's earliest study was two-dimensional~\cite{pager_mark_2003}. The study worked with all-male testers, varying tester race and whether the CV included a criminal record. Had Pager added gender, the study would require three dimensions. Had Pager considered additional variations in CV details such as education level, the study would have grown further in complexity. Increased dimensional complexity creates challenges of recruitment, training, and coordination.

%The number of testers increases exponentially with each new variable and linearly with each new possible attribute for a given variable. %In our audit of political advertising policies, if we had audited binary categories of U.S./non-U.S./location, bank account location, U.S. currency, and citizenship status, we would have a $4^2$ matrix of identity characteristics. 
%As more variables are added, the complexity of a study could quickly become too large for humans to coordinate alone.

%Among the prompts being tested, dimensional complexity creates challenges of design and coordination. 

%For example, auditing political advertising policies is dimensionally-complex in a country with two major parties, 50 states, and a range of issues where companies could make errors in the enforcement of their policies. When auditing election advertising policies, testers might need to prompt companies with many different prompts which are somehow similar enough for comparison, different enough to avoid detection of the audit itself, and tailored to the specific variables being tested. The effort of designing these prompts and allocating them to testers would quickly become too complex for humans to conduct efficiently.

\subsubsection{Trading off Dimensional Complexity and External Validity}
Since findings from audit studies only generalize to the kind of testers and prompts being sampled, audits face a trade-off between complexity and external validity. In the social sciences, external validity refers to the repeatability of a study outside the lab and also to the degree to which findings describe phenomena that actually occur in the field \cite{shadish_experimental_2002, cialdini_full-cycle_1980}. Simple audits that test a small number variables can detect decision-making errors that are important to society and to science. Yet even when researchers match testers and prompts to common cases, the findings cannot describe the performance of a decision-making system outside the audit sample. For example, while Pager's study established discrimination among male job applicants, it took other audits to establish systematic hiring discrimination toward women \cite{bertrand_are_2004}.

To put it simply, designers of audits face a trade-off between external validity and dimensional complexity. A more complex study with more variables can include a wider range of people and prompts. A narrower study with fewer variables will cover a more constrained range of cases. This trade-off has high-stakes consequences when auditing complex decision-making systems, such as the ones we study in this work. In such high-dimensional settings, auditors could miss important errors because the test did not cover a wide enough range of prompts.

While increased complexity improves the breadth and depth of a study, all audit studies face common limitations in generalizability. To discover if errors are persistent over time, researchers typically publish multiple individual audit studies which are then later meta-analyzed~\cite{quillian_meta-analysis_2017, ross_measuring_2017}. In computer science, since decision-making software can be easily changed by its creators, audit studies may never discover patterns that persist over time \cite{kitchin_thinking_2017}. In those circumstances, low-cost, high-validity audit studies become an even more important tool for evaluating and altering decision-making systems \cite{mullainathan_biased_2019}.

% remove this sentence?
By creating software to automatically generate audit materials, estimate sample sizes, allocate prompts across testers, and generate statistical results and illustrations, we hoped to increase the validity of audit studies by expanding their dimensional complexity.

\subsection{Software-Supported Audit Studies}
Computer scientists have developed systems for high-dimensional audits of decision-making systems in the lab \cite{adler_auditing_2018, kearns_preventing_2018}. Yet the work of conducting audits in the field is constrained by the challenges of working with testers and designing prompts for those testers to use. %While computer scientists have made progress on the recruitment and coordination of testers, the generation of prompts has remained a manual process.

What does software for audit studies look like? Scientific software often exists as a collection of libraries, scripts, and best practices that help researchers automate various parts of their work~\cite{wilson2014best}. For example, the OpenWPM system, which has supported many observational studies of websites, began as a collection of scripts that have supported many influential studies on based on web-scraping~\cite{englehardt_online_2016}.

In some domains, software tools have coordinated large-scale volunteer monitoring of decision-making systems that shape network conditions, such as censorship and filtering. Two prominent examples are OONI \cite{filasto_ooni:_2012} and Encore \cite{burnett_encore:_2015}. In the case of OONI, volunteers run software on their own machines to test various aspects of network connectivity, such as the reachability of various Internet destinations. Encore can recruit and coordinate a much larger user base, since it does not require special software installation, and since tests can occur with minimal to no training.

Computer scientists have simulated human testers with software. By removing humans from the research process, scientists can generate testers with a wide range of characteristics \cite{tschantz_methodology_2015}. Despite the efficiency and variety of this promising approach, questions remain about the reliability of a method that could be interpreted by decision-making systems as inauthentic or fraudulent activity~\cite{narayanan_princeton_2017}.

Social scientists have created externally-valid audit prompts by manually adapting real-world content. In one audit of employment discrimination, researchers downloaded resum\'es from an online website and edited them to create their decision prompts \cite{bertrand_are_2004}. In a study of online censorship in China, researchers looked for news articles about political and non-political collective action that they then tested posting to Chinese social media \cite{king_reverse-engineering_2014}. While drawing from real cases increases the naturalism of an audit study, the authors of these studies use labored words like ``scour'' to describe the substantial effort required  by this manual process. Neither study considered more than two binary variables, likely constrained by the substantial effort of finding examples and adapting them into scientifically-valid prompts.

\section{Design Considerations for Software-Supported Audits of Decision-Making Systems}
Any designer of software to support audit studies needs to consider five basic factors in their system design: variable selection, audit prompt creation, statistical power, volunteer coordination, and law/ethics.

\subsection{Choosing Variables to Test}
With any audit study, researchers must decide which characteristics of the testers and the characteristics of the prompts that they wish to test and which ones they wish to hold constant. These decisions determine the dimensional complexity of the study and are constrained by the feasible sample size required to observe meaningful error rates. While larger, more complex audits offer greater precision and validity, they also require more testers, more time from testers, and more money. Study designers must also define which variables can be altered in the content of prompts (such as political leaning or educational level) and which ones rely on participant recruitment (such as skin tone or nationality) \cite{sen_race_2016}. 

%Audit studies also vary the prompt provided to the decision-making system being tested. In the case of political advertising policies, we were interested in democrat and republican candidates, progressive and conservative issues, and elections at a federal and local level.

 %Researchers must choose how many variations to include in the study design. In the audit of political advertising policies, we considered the citizenship of the tester, the geolocation of their internet activity, browser and platform language settings, whether their bank account was based in the U.S., their billing address, and what currency their bank account used.

%To support the design of our audit study, we wrote software that took input on the characteristics and prompts we wished to test. The software bundled those characteristics into combinations of personas and prompts, and simulated the sample size needed to observe meaningful differences in error rates. 

\subsection{Generating Realistic Audit Prompts}
In any audit study, the validity of the study depends on the realism of the prompts--the occasions and materials that force the decisions recorded by testers. When auto-generating prompts, researchers risk losing that validity in favor of automation. Yet automation can also improve the realism of audit studies when researchers draw from a range of qualitative and quantitative evidence, including news stories about platform mistakes, public data about an election, and other relevant public datasets.

\subsection{Design Diagnosis and Statistical Power}\label{sec:statistical_power}
Since audit studies set out to describe systematic errors through statistics, researchers need to specify a sample size for the number of testers and the number of prompts. Deciding on a sample size is often an iterative process of making decisions in light of trade-offs of complexity, precision, and cost. With larger samples, researchers can estimate base rates and error rates with greater precision. As dimensional complexity increases, researchers must either increase the sample size or reduce the precision of their findings. If the sample size is too small, researchers may fail to observe important forms of systematic error. Yet researchers are also motivated to limit sample sizes since larger samples require more time, coordination effort, and direct costs.%, researchers also have reasons to seek the smallest sample size that matches their precision goals.

% For low-dimensional studies (such as an audit study based on one binary variable), researchers can use simple power analysis tools to choose a sample size. In high-dimensional studies, researchers may wish to observe error at different levels of precision for different characteristics. 

%For example, in an audit of political advertising policies, differences in error rates based on currency type might be on a different scale or have a different importance from errors that differ by political party. Researchers may also want to allocate sample sizes to observe party-based errors with greater precision. Researchers may also have a limited number of testers or seek to operate within certain financial constraints. Commonly-available power calculators cannot support such levels of dimensional complexity.

In the social sciences, researchers explore these trade-offs in a process that examines the Model, Inquiry, Data Strategy, and Answer Strategy (MIDA) of a study \cite{blair_declaring_2019}. In this approach, researchers create a Model (M) by describing and simulating the processes they wish to be able to observe in the field---in our case, the decision-making system. Researchers then also simulate their data collection process (Inquiry, Data), which may also have sources of error. Finally, researchers diagnose the details of their study, from sample size to analysis plan (Answer), by examining the results of the full simulation.

% Audit studies must also conduct a power analysis to determine how many prompts should be generated and placed to arrive at results with statistical confidence. Common software packages can help researchers conduct power analyses with each variable that they are interested in testing. Depending on the dimensional complexity of the study, the resources available to researchers, and the degree to which the study can be automated, researchers may need to make a trade-off between statistical confidence and the number of audit prompts that are created. In any case, an exploratory power analysis with software running many different simulations of an audit study can help researchers empirically understand this trade-off.

\subsection{Involving and Coordinating Volunteers}
The selection, training, and coordination of testers are essential to any audit study. By providing platforms for targeted advertising, companies have simplified the recruitment of participants into audit studies \cite{matias_who_2017}. Software can also play a key role in coordinating participants, especially when audit studies require large-scale and personalized coordination \cite{burnett_encore:_2015}. 

%When designing this audit, we considered using Facebook and Google's own advertising targeting systems to recruit volunteers with the required characteristics, as other algorithm audits have done \cite{matias_who_2017}. We also considered creating software to automatically place ads on behalf of testers, reducing the effort and training required of testers. 

%Since the study as we designed it required a small number of testers, we did not implement these ideas for recruiting and coordinating participants. We did however design the prompt-generation software to allocate them to testers in correspondence with the study design.

Researchers conducting audit studies of online platform policies should also consider the risk of companies observing communications between the researchers and among testers. Maintaining secrecy can be difficult when testing platforms that also monitor communications between individuals. Designers of software for audit studies may need to consider end-to-end encryption when developing systems.

% group chat software from an organization independent from Facebook or Google.\footnote{\url{https://element.io}}

\subsection{Law and Ethics}
Researchers who conduct audit studies must always consider the legal and ethical risks for four parties: the testers, people who might encounter the prompts, the parties that oversee the decision-making system, and the researchers \cite{pager_use_2007}. Researchers face legal risks if the audit could be interpreted as a violation of the law. Testers could also face risks if their behavior was interpreted as illegal or if conducting the audit might introduce irreparable harms to how they are treated by the organization being tested. Members of the public could also be harmed by the audit process itself, for example if their job prospects or political behavior were influenced by a study. Finally, audit studies can create risks for people within the organizations being tested, especially if their employers hold them responsible for systemic errors revealed by a study.

Careful legal and ethical consideration are especially important with audit studies, which sometimes fall outside of the scope of university ethics boards. Our own institution's IRB determined decided that our audit it does not fall under the purview of the U.S. Common Rule for research ethics, since in their view it constitutes a contribution to institutional rather than generalizable knowledge. In their view, research about Facebook and Google's decision-making systems at a moment in time was not generalizable enough for them to review. Despite this decision, we have followed standard practices in academic research ethics to minimize the risk from our research and to protect the privacy of those involved--seeking consent from testers, storing all data securely, anonymizing the data, and minimizing the number of people who were exposed to the ads.

\section{Software-Supported Auditing in Practice in the 2018 U.S. Midterm Elections}
In 2018, we developed software to support an audit of Google and Facebook's political advertising policy enforcement systems. We first provide background on the political advertising policies of Facebook and Google in 2018. We then present the design and results of the audit, followed by what researchers can learn from this experience about software-supported audit studies.

%We recruited a team of 7 co-investigators to help post 477 ads before the 2018 U.S. midterms and analyze the companies' policies.
%We found systematic errors in how the companies enforced their policies, but we found no evidence of partisan bias.

\subsection{Platform Political Advertising Policies in 2018}
%In this paper, we test a software-supported approach to audit studies in an audit of Facebook and Google's political advertising policies during the 2018 US midterm elections.

Since the 2016 United States presidential election, voters have become more aware of the potential for online advertising on social media platforms to influence elections.
For example, Russian actors placed election-related ads on Facebook to influence American voters, which covered topics such as gun control, racial tension, and immigration~\cite{politico2018-indictment, facebook2018-questions}.
Politicians criticized Facebook for allowing these ads to be published \cite{cspan2017facebook}. Many insisted that Facebook should have detected and prevented these attempted influence campaigns by observing available signals such as payment currency--the Russian ads were paid for in rubles~\cite{hill2017rubles}.

In response to public concern and U.S. Congressional hearings, Facebook \cite{madrigal_will_2018} and Google \cite{reardon_politicians_2018} implemented policies in 2018 to limit who can publish election-related ads during national elections. During several national elections, including the U.S. 2018 midterm election, companies required advertisers to confirm their identity and nationality before permitting them to publish ads promoting candidates in the period before the election. By doing so, companies were forcing advertisers to comply with U.S. Federal Election Commission (FEC) regulations that required disclosure of who funded federal election advertising campaigns \cite{valentino-devries_i_2018}.

Facebook also developed policies that restricted who could publish so-called ``issue ads,'' a category that regulators have struggled to define. In the U.S. legal tradition, issue ads include content that a ``reasonable person'' might guess are about an election. Issue ads evade election regulations by avoiding direct encouragement to vote for a candidate \cite{hayward_when_1999}. These ads are common and controversial in U.S. elections. Issue ads are also the kind of ads published by foreign influence operations during the 2016 U.S. presidential election. To simplify decades of policy making and Supreme Court cases, because issue ads are to difficult to define clearly, the FEC did not restrict issue ads or require funding disclosure for these ads in the 2018 election \cite{hayward_when_1999, valentino-devries_i_2018}. Under pressure from politicians to do what the government considered too complex, Facebook developed policies in 2018 that forbade advertisers from publishing ads about ``issues of national importance'' without verifying themselves.

\subsubsection{Prior Research on Political Advertising Markets}
Researchers have studied these political advertising policies using approaches from information security and algorithm design.

Security researchers have investigated how the Russian Intelligence Research Agency (IRA) used Facebook's ad targeting tools to create politically divisive ads before the 2016 election~\cite{ribeiro2019microtargeting}. 
%They found that Facebook's data collection process for advertising enabled the Russian actors to target specific populations of users that might be susceptible to political influence. 
Other security researchers have identified ways that malicious advertisers could prevent their political ads from being disclosed in Facebook's political ad library~\cite{edelson2020security}. 

%User experience researchers have studied the incomplete and misleading nature of platform explanations for why a user has received an ad~\citep{andreou2018investigating}.

Since machine learning models inform the decision to label content as political advertising, researchers have studied platform policies from the perspective of algorithm and market design. One team developed a browser plugin to gather Facebook ads shown on volunteers' timelines. After using machine learning to classify ads as political or not, they found cases of political ads that Facebook failed to label~\citep{silva2020facebook}. Another team showed that the content of an ad on Facebook (e.g., its associated image) can cause an ad to be disproportionately delivered to one group of users over another, despite the same targeting criteria and the same bidding strategy being used across ads~\cite{ali2019discrimination, ali2019ad}.

\subsubsection{Decision-Making Systems for Political Ads}
To determine which advertisers needed to be authorized, Facebook and Google created decision-making systems that used a combination of machine learning and human processes to review whether a given ad would be required to comply with a company's political ad policies. In 2017, Facebook announced that they would hire 1,000 content moderators to help review ads. They also developed a machine learning system to detect "inauthentic Pages and the ads they run" \cite{techcrunch2017facebook}. While Google was less open about their methods, they also reviewed ads for political content and required advertisers for political candidates to verify their identity~\cite{reardon_politicians_2018}.
% Given that Facebook reviews \textit{millions} of ads each week, their systems are bound to falsely block non-election ads~\cite{facebook2017-operations}.

% Austin - can you include more information about specifically what the rules were for Facebook and for Google, one sentence for each platform? Mention how many people Facebook said it was hiring to handle content moderation ^  

%We may want to publish a diagram of the process, which is roughly like this:

% 1. Attempt to publish election-related ad
% 2. The company reviews the ad before or after the ad is published
% 3. After review (sometimes after publication), the ad is prevented and the verification process is required

Upon identifying and verifying an advertiser, both Facebook and Google promised to disclose the advertiser's identity when publishing the ad, as well as publish reports and datasets of election-related advertisers~\cite{bowles_facebook_2018}. Both companies published interactive websites that provided information about the entities who had published past and present political advertisements~\cite{google2018transparency, techcrunch2017facebook}. In their press releases, both companies expressed hope that these datasets would provide greater transparency toward advertisers as well as disclosure of the decisions made by companies \cite{google2018transparency, leathern_shining_2018}.

\subsubsection{Systematic Errors from Political Ad Policy Enforcement}
Google and Facebook's political advertising policy systems, like any decision-making system, can make systematic errors. In previous decades, the U.S. government had restrained political advertising regulations to protect freedom of expression, since mistakes could substantially impact both an election and American civic life \cite{hayward_when_1999}. Systematic errors by platform enforcement systems introduce three risks to society:
\begin{itemize}
    \item Platforms might be \textit{too permissive}, creating systems that were ineffective at preventing, removing, or labeling ads that violate a platform's policies and election law.
    \item Platform decision-making systems might be \textit{too restrictive}, forbidding important public discourse protected by the constitution that was unrelated to elections and permitted by a company's own policies.
    \item Platforms could also be \textit{politically-biased} if systematic errors advantaged or disadvantaged a given political candidate, party, or ideology.
\end{itemize}

Throughout 2018, platforms were accused of all three kinds of decision-making failures. Google's decision-making system was accused of being too permissive by failing to apply its political advertising restrictions to ads by major political candidates \cite{wsj2018google}. Facebook's decision-making system were accused of being politically biased by failing to label a political ad that asked San Francisco voters to vote "yes" on a school bond proposition and by blocking non-partisan ads from newspapers \cite{propublica2018screening}.  Lastly, Facebook was accused of being too restrictive by falsely labeling ads for veterans, LGBTQ+ people, and even baked beans whose brand was similar to the name of a former U.S. president \cite{gale_facebooks_2018, rosenberg_facebook_2018, mak_facebook_2018}. All of these cases involved policy actions that were inconsistent with a platform's own written policies.

\subsubsection{How Audit Studies Advance Understanding of Political Advertising Policies}
While stories about political ad policy enforcement errors were observed in platform transparency reports, none of the datasets published by Facebook or Google during the 2018 election can be used to ask if the policies were too permissive, too restrictive, or politically biased. To answer these questions, researchers would need to compare restricted accounts and ads to those that were permitted. Since transparency datasets only include ads that companies labeled as election-related, it is impossible to make this comparison. Even if platforms were to publish the full set of all ads submitted during the election period, observational studies could still produce an inaccurate picture of the fairness of a company's processes. For example, if one candidate in an election was more or less competent at online advertising than their opponent, observational transparency data might lead researchers to mistakenly conclude that the company's decision-making system wrongfully favored the more competent advertiser.

Unlike correlation studies of administrative data, audit studies evaluate decision-making systems under balanced and controlled conditions in the field. Consequently, audit studies provide reliable tests of a system's average performance, including any cases of under-enforcement, over-enforcement, and bias. Yet audit studies have typically been too difficult to design and coordinate for complex processes like elections, which involve many candidates campaigning across many regions with local issues.

\subsection{Research Question: Auditing Mistaken Advertising Policy Enforcement by Facebook and Google}
What \textit{kinds of ads} do Facebook and Google mistakenly prohibit? Furthermore, \textit{what percentage} of non-election ads are wrongly prohibited by these companies? By asking these questions, our audit study examines the problem of over-reaches in platform content policy enforcement, an issue that scholars of human rights have identified as a major risk to civil liberties worldwide \cite{seltzer_free_2010, mackinnon_consent_2012}. Because freedom of expression and assembly are internationally-recognized human rights, the question of mistaken enforcement is important to society independently of harms that companies mistakenly permit.

By investigating the rate at which non-election ads are prohibited, we are studying the decisions of a platform's enforcement systems on average. Policy enforcement mistakes could result from many factors, including the details of a company's policies, the quality of training, the behavior of automated filter software, and perhaps differences in the judgment of individual workers. Because audit studies cannot typically distinguish between internal organizational factors, and because we evaluate the system as a whole rather than individuals, these findings should never be interpreted as a reflection upon any individual worker enacting policy for these companies.

We carefully note that we do not study whether Facebook and Google are too permissive in their enforcement of their political advertising policies. However, to study this question, we would need to place campaign-related ads, which introduces significant legal and ethical risks for multiple parties. For example, the lead researchers on this study are employed by a university that holds a 501(c)(3) non-profit status. According to U.S. federal law, 501(c)(3)s are forbidden from participating in electioneering, which includes placing advertisements that advocate for a candidate. Testers could also face risks if their behavior was interpreted as illegal or if conducting the audit might introduce irreparable harms to how their user accounts are treated by a company~\cite{desposato_ethical_2014}. Members of the public might also be shown ads  that influence their political beliefs and behaviors. Lastly, systematic errors might be wrongly attributed to the individuals making decisions, who are often in precarious employment positions~\cite{roberts_behind_2014}.

\subsection{Software-Generated Audit Prompts}
To answer our research questions, we created software to support an audit of decision-making systems for political advertising policies pertaining to the United States. Our software queried publicly-available APIs to help generate ads that testers published to targeted geographically-targeted audiences on each platform. Our software also simulated a power analysis to determine how many ads needed to be placed for each tester, and it allocated ads to their respective testers. We also wrote code to generate statistical results and illustrations. Each observation combined three high-level variables: Ad Type, Leaning, and Location. We summarize the methods of our audit study below.

\begin{figure*}[h!]
  \centering
  \begin{subfigure}[b]{0.23\textwidth}
    \includegraphics[width=\textwidth]{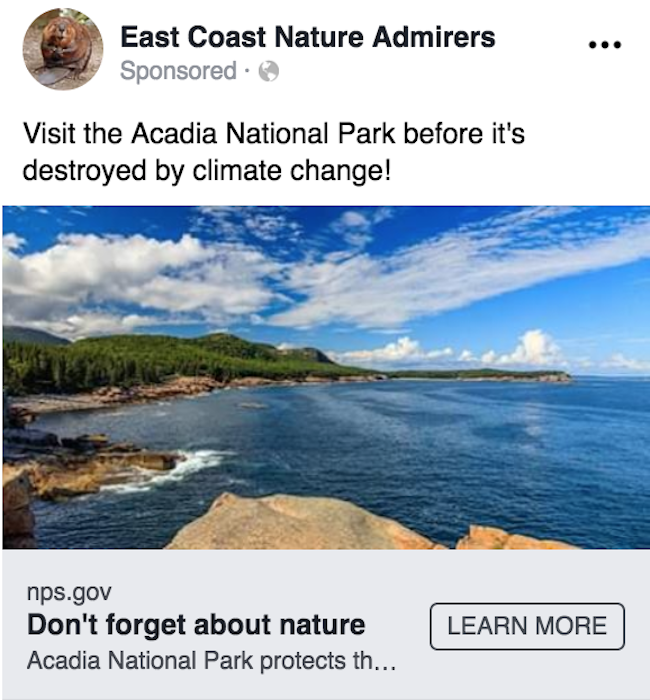}
    \caption{Left-leaning parks/parades ad}
    \label{fig:facebook_ad_1}
  \end{subfigure}
  \begin{subfigure}[b]{0.23\textwidth}
    \includegraphics[width=\textwidth]{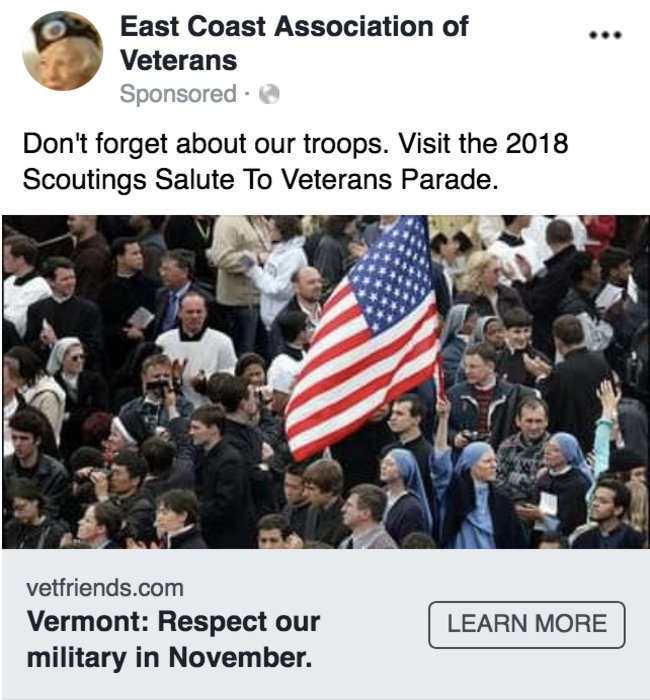}
    \caption{Right-leaning parks/parades ad}
    \label{fig:facebook_ad_2}
  \end{subfigure}
  \begin{subfigure}[b]{0.23\textwidth}
    \includegraphics[width=\textwidth]{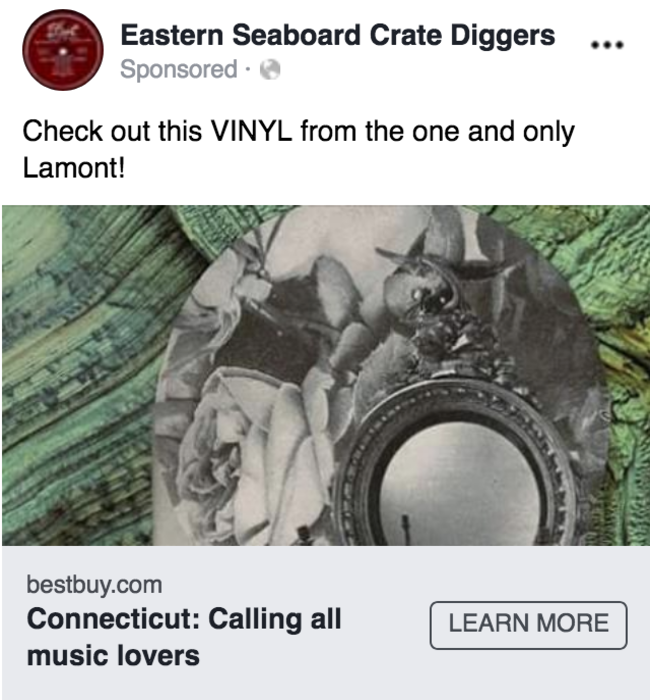}
    \caption{Left-leaning product ad}
    \label{fig:facebook_ad_3}
  \end{subfigure}
  \begin{subfigure}[b]{0.23\textwidth}
    \includegraphics[width=\textwidth]{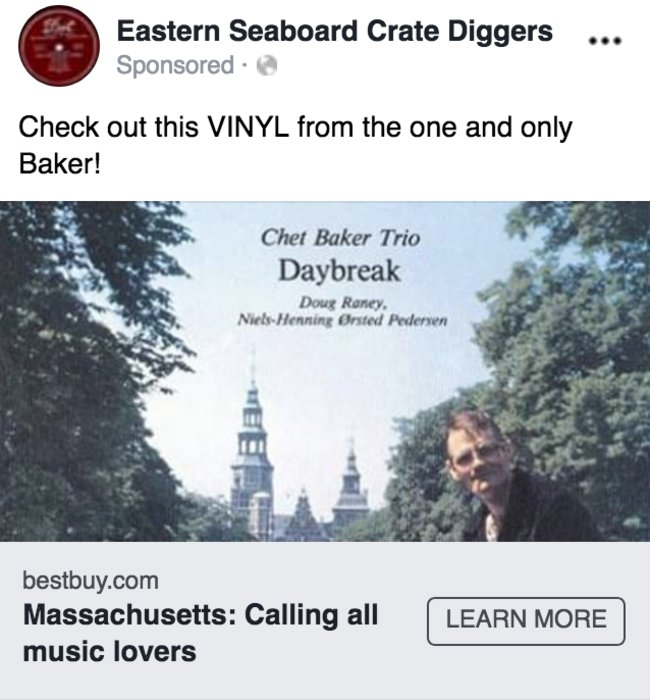}
    \caption{Right-leaning product ad}
    \label{fig:facebook_ad_4}
  \end{subfigure}
    \caption{Previews of Facebook ads that were posted to one of our Facebook pages during this study}
  \label{fig:facebook_ad_examples}
\end{figure*}

\begin{figure*}[h!]
  \centering
  \begin{subfigure}[b]{0.48\textwidth}
    \includegraphics[width=\textwidth]{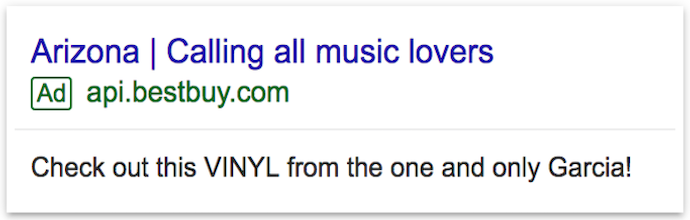}
    \caption{Left-leaning product ad}
    \label{fig:google_ad_1}
  \end{subfigure}
  \begin{subfigure}[b]{0.48\textwidth}
    \includegraphics[width=\textwidth]{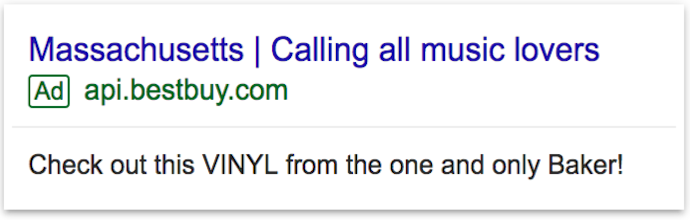}
    \caption{Right-leaning product ad}
    \label{fig:google_ad_2}
  \end{subfigure}
    \caption{Previews of Google ads that were posted during this study.}
  \label{fig:google_ad_examples}
\end{figure*}

\subsubsection{Designing Relevant Ads}
We designed three kinds of ads in consultation with multiple U.S. election lawyers: product mistakes, community event mistakes, and government website mistakes. To test erroneous removal of non-election product ads, we chose products with names that included the surnames of political candidates, based on reports that Facebook required authorization by the makers of ``Bush's Beans", an American food company that shares its name with former United States presidents \cite{rosenberg_facebook_2018}.

To test erroneous removal of ads for community events, we included ads for Veterans Day celebrations after Facebook reportedly removed ads for non-election gatherings of LGBTQ people and websites for U.S. military veterans on the grounds that they were election-related \cite{mak_facebook_2018, gale_facebooks_2018}. In these cases, Facebook mistakenly prevented non-election advertisements from being published, and the company has apologized and changed its decisions after news articles mentioned those decisions.

Finally, to test erroneous removal of ads linking to government websites, we designed ads for national parks after hearing informal reports, later corroborated, that Facebook was preventing some government services from publishing ads about non-partisan public service information \cite{smith_facebook_2019}. Since platforms could legitimately block advertisements that contain links to government websites, we chose websites that focus on upcoming events and places rather than news. We associated a rightward political leaning to Veterans Day ads emphasizing the respect of the military and a leftward political leaning to environment-focused ads encouraging use of public parks.

% In the past, researchers have created materials for audit studies by hand, which limits the  dimensional complexity of the studies and their sample size \cite{bertrand_are_2004}.
% By writing custom software to do this work for us, we were able to conduct a large-scale, in-depth study to understand the complexity of Facebook and Google's decision-making systems.
We summarize our tested variables below:

\begin{itemize}
    \item \textbf{Ad type}: Whether a non-political ad could be mistaken for:
    \begin{itemize}
      \item Candidate support ad 
      \begin{itemize}
        \item Product ads for music albums in which the artist shared the last name of a political candidate for the 2018 U.S. Congressional midterms and gubernatorial elections.
      \end{itemize}
      \item Issue ad
      \begin{itemize}
        \item National park ads that encouraged people to visit a particular park before it is "destroyed by climate change". 
        \item Veterans Day parade ads that encouraged people to visit a particular parade, "respect the military", and to "remember the troops". 
      \end{itemize}
    \end{itemize}
    \item \textbf{Leaning}: whether the ad could be mistaken for left or right leaning content, or for supporting Republican or Democrat candidates
    \item \textbf{Location}: the targeted location of the ad, based on a specific election that the ad could be mistaken for: 
    \begin{itemize}
        \item A state (governor) or federal (house) election
        \item Geographically-targeted advertising toward regions voting in that election, either a state (governor) or voting district (house)
    \end{itemize}
\end{itemize}

\subsubsection{Ad Generation}
Once we designed types of ads to place and determined how many ads should be placed, we wrote software to help generate ads for Facebook and Google. Figure~\ref{fig:facebook_ad_examples} and Figure~\ref{fig:google_ad_examples} show examples of our ads. 

To generate non-election product ads, our software scraped data about 2018 U.S. Congressional midterm elections from the Federal Elections Commission. The FEC API provided up-to-date details on every candidate running for federal office, which party they were from, and which voting districts are voting on which positions.\footnote{\url{https://api.open.fec.gov/developers/}} We also manually scraped the names of candidates for gubernatorial elections from Wikipedia.\footnote{\url{https://en.wikipedia.org/wiki/2018_United_States_gubernatorial_elections}} Our software then queried the BestBuy product API for music albums that shared a surname with a candidate.\footnote{\url{https://bestbuyapis.github.io/api-documentation/}} For the image of each ad, our software scraped the image of the first music album that was returned for a query with a candidate's surname. For the body text of each ad, our software used a candidate's surname and the physical format of the respective music album to generate text that read "Check out this <Album format> from the one and only <Candidate surname>!" For the header text of each ad, our software used a candidate's surname and the physical format of the respective music album to generate text that read "[State]: Calling all music lovers."

To generate ads for government websites, our software scraped the list of U.S. National Parks, the park location in relation to voting districts, and the website for each park from an API provided by the National Park Service.\footnote{\url{https://www.nps.gov/subjects/digital/nps-data-api.htm}} For the image of each ad, our software scraped the image returned by the API for the national park. For the body text of each ad, our software used the name of a national park to generate text that read "Visit the <National park> before it's destroyed by climate change!" For the header text of each ad, we wrote "Don't forget about nature."

Finally, to generate ads for community events, our software scraped a list of local observances of Veterans Day--an non-partisan national holiday held on November 11th, 3 days after election day--from \url{vetfriends.com}. At the time we scraped the data, the website listed the details of Veterans Day parades across the United States. For the image of each ad, we manually accessed images from a collection of Wikimedia Commons images of the U.S. flag. Our prompts matched the location of these celebrations with voting districts in the test. For the body text of each ad, our software used the name of a Veterans Day parade to generate text that read "Don't forget about our troops. Visit the <Parade name>." For the header text of each ad, our software used the respective state that a parade was planned for to generate text that read "[State]: Respect our military in November."

\subsubsection{Ad Placement}
To place our ads, we recruited  7 testers who are U.S. citizens and who have characteristics that we thought might influence the chance of an advertisement to receive enforcement. We allocated our advertisements to testers by sampling each combination of ad type, investigator type, and location. Where the platform required ads to be associated with a group, channel, or page, investigators created a separate page for each ad type. 

The two types of testers were:

\begin{itemize}
    \item US: U.S. citizens with an EN-US browser locales and U.S. IP address locations using U.S. Dollars to place ads
    \item Non-US: U.S. citizen with a non-US browser locale and non-US IP address, using a non-US bank and non-US currency (CAD, GBP) to place ads
\end{itemize}

Each tester created an advertiser account on Google and Facebook, as well as a Facebook page for every ad type that they tested. Pages offered geographically-specific content related to the topic of the ads that were associated with them. Testers posted Adwords ads to Google on auto-generated search terms relevant to a given ad. Testers attempted to publish each ad for a period of 48 hours at a budget of 1 unit of currency per day (US, CAD, GBP). Testers then recorded whether the ad was published or prevented from being published by the platform, citing their political advertising policies. 

Since none of the prohibited ads were published by the platforms and since none of the published ads were publicly labeled by platforms as election-related, none of these ads appeared in platform transparency reports. Although we had intended testers to seek authorization and record how platforms reported the identity of advertisers, we realized during the audit that appearing in platform transparency reports would de-anonymize our testers and represent an undue burden.

\subsubsection{Software-supported Research Design Diagnosis}
To refine the final study design, we wrote power analysis software that implements the MIDA process for design diagnosis of audit studies (as described in Section \ref{sec:statistical_power}). First, we developed inputs on the minimum observable difference and bias we wanted to be able to observe for different combinations of personas and ads. Using that information, the software simulated thousands of possible audit studies to estimate the number of prompts and volunteers that would be needed. For each possible sample size and data collection plan we considered, this configurable software simulated estimates and error bars for multi-variate comparisons on each of our tested variables--ad type, political leaning, targeting location of the ad, and the location of the testers (U.S. or non-U.S.) (Appendix Figures 1, 2, and 3).

%% EDITED
To choose the target statistical power for our audit study, we considered the pragmatic impacts of systematic errors. For a single nonprofit organizing a community celebration, an inability to advertise could reduce turnout and donations. Yet we were also aware that large-sample studies can also reveal small differences that are statistically significant if not pragmatically meaningful~\cite{morin_facebooks_2014}. Given how little influence online ads exert on American political behavior~\cite{coppock_small_2020} and how few ads received clicks in the quarter before the election~\cite{noauthor_q2_2018}, we designed our study to detect differences that would be undeniably meaningful if observed. Assumed that base rates would be close to 100\% ads published, we then aimed to observe differences of at least 5 percentage points for our top-level analyses. With $\sim$20 ads for each combination of a platform, investigator location, election location, political leaning, and ad type, each of our main analyses would have a sample of 240 prompts.

    % \begin{figure}[h!]
    %     \includegraphics[width=\textwidth]{figures/simulation-estimated-rates-n-50.png}
    %     \caption{Simulated chance of publication for a given ad combination (persona, election, advert). Hypothetical ``true'' block rates are indicated with an orange X. Total ad placements: 1800. Ad placements per combination:50.}
    %     \label{fig:poweranalysis}
    % \end{figure}

%Pre-registration is becoming the norm for demonstrating research integrity in many social science fields~\cite{lindsay_preregistration_2018}. The variables we considered include:

% \begin{figure*}[h]
%     \centering
%     \includegraphics[width=0.75\textwidth]{figures/examples-of-non-election-ads-facebook.png}
%     \caption{Previews of Facebook ads that were posted to one of our Facebook pages during this study (actual ads may have been posted to a different page).}
%     \label{fig:facebook_ad_examples}
% \end{figure*}
% 
% \begin{figure*}[h]
%     \centering
%     \includegraphics[width=0.9\textwidth]{figures/examples-of-non-election-ads-google.png}
%     \caption{Previews of Google ads that we posted during this study.}
%     \label{fig:google_ad_examples}
% \end{figure*}

\subsubsection{Analysis Plan}
We pre-registered the analysis and results-generating code at the Open Science Framework before collecting any data.\footnote{\url{https://osf.io/4zudh}}
All of our code, data, and training materials are publicly available on GitHub.
We have published these materials to be completely transparent about how we carried out this study.\footnote{\url{https://github.com/citp/mistaken-ad-enforcement}}

\subsection{Audit Study Results}
From 2018-09-17 through 2018-10-10, our team of 7 posted a total of 477 ads to Facebook and Google.\footnote{One intended Facebook ad was not found in our final records and may not have been posted. We have removed this observation from the analysis} We observed whether the ad was prevented from being published by the platform for allegedly violating policies about election advertising.\footnote{One product ad was blocked by Facebook because the platform judged that the cover image included too much skin. We manually chose a different image and made another attempt, which was published by the platform. We did not count this as an ad blocked for its relation to the election.}

Google did not prohibit any of the 239 ads that we posted to their platform. Facebook however prevented 10 out of 238 ads from publication, citing their election policies, a total of 4.2\% of the ads we placed.\footnote{To confirm that these ads were genuinely permissible, we submitted two of the ads to Facebook's appeals process, and the company reversed their decision for both.} Parks and parade ads were 9 of the ads that Facebook prohibited, and 1 of them was a product ad. Among ads that Facebook prevented from publishing, 3 could have been mistaken for being right leaning or for Republican candidates, and 7 might have been mistaken for being left leaning or for Democrat candidates. 

\begin{figure*}[h]
    \centering
    \includegraphics[width=0.99\textwidth]{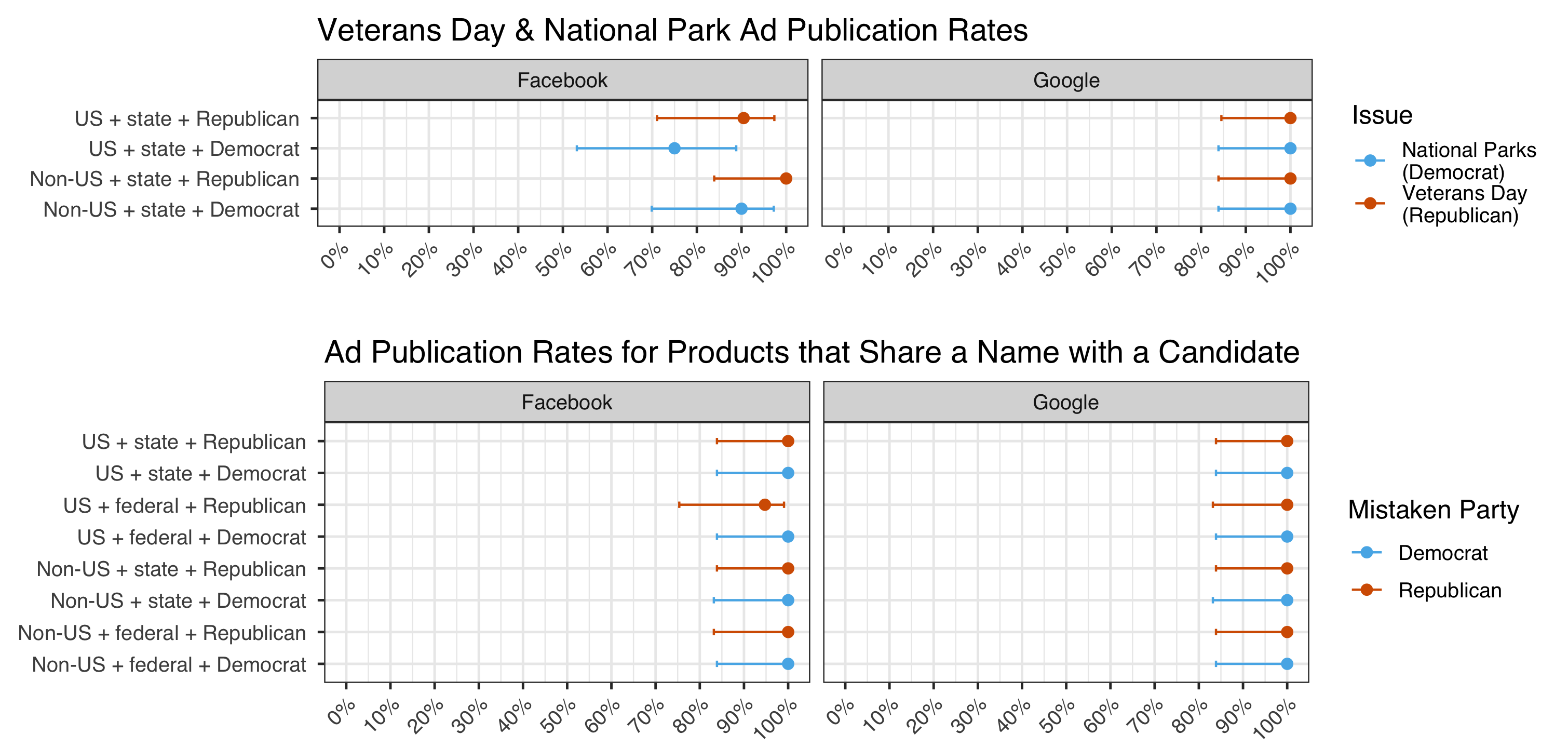}
    \caption{Estimated chance of publication for a given ad combination (election, political leaning). 477 ad placements were attempted by 7 people from 2018-09-17 to 2018-10-10. Product ads are music albums that share the artist's last name with a candidate. Veterans Day \& National Park ads are about events and places that could be mistaken by platform policy enforcers as election-related ads of national importance. 95\% confidence intervals use the Wilson method. Code \& data: \url{https://github.com/citp/mistaken-ad-enforcement}.}
    \label{fig:per_characteristic_results}
\end{figure*}

Our main analysis estimates the rate at which a certain type of advertisement is permitted. To do so, we computed the groupwise means and confidence intervals for the chance of an ad of a certain kind posted by a certain kind of person to be permitted by a platform. Groupwise means and confidence intervals are generated using the Wilson estimation method for confidence intervals (the binom.confint function in the R library \textit{binom}). At small sample sizes, a small increase in the number of ads placed may lead to large differences in the calculated 95\% confidence intervals. The Wilson estimation method minimizes variation in confidence intervals between small differences in the sample size at smaller samples \cite{brown_interval_2001}. Groupwise means and confidence intervals for each tested characteristic are available in Figure \ref{fig:per_characteristic_results} and Table 2.

\begin{figure*}[h]
    \centering
    \includegraphics[width=0.99\textwidth]{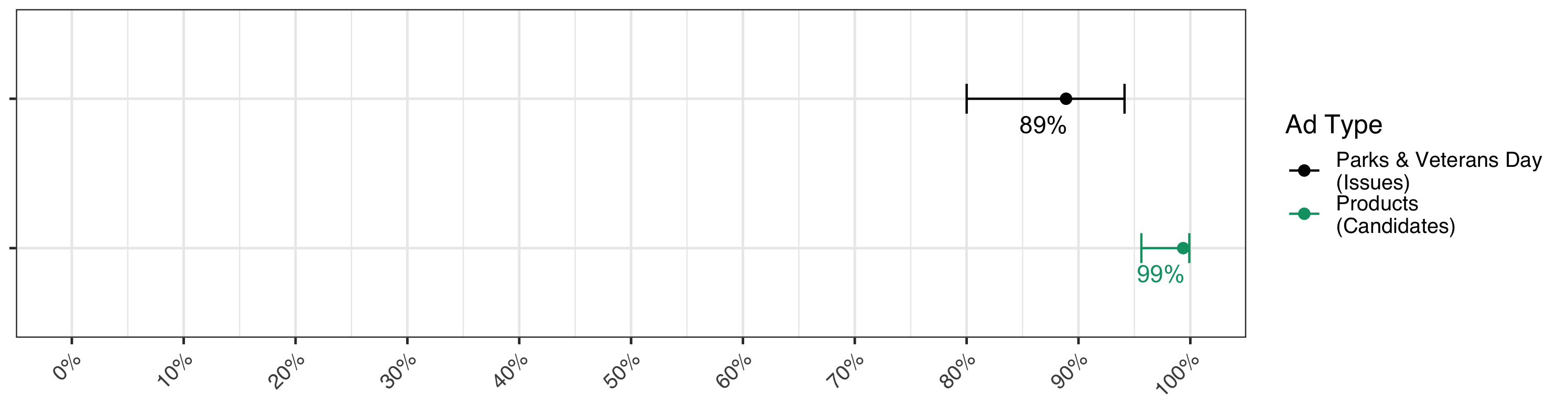}
    \caption{Estimated Facebook publication rate for non-election advertisements comparing Veterans Day \& National Park ads (issue mistake) to Product ads (candidate mistake). 238 ad placements on Facebook were attempted by 6 people from 2018-09-17 to 2018-10-10. Product ads are music albums that share the artist's last name with a gubernatorial candidate. Veterans Day \& National Park ads are about events \& places that could be mistaken by platform policy enforcers as election-related ads of national importance. Results from a logistic regression (p=0.005). Code \& data: \url{https://github.com/citp/mistaken-ad-enforcement}.}
    \label{fig:issue_vs_candidate}
\end{figure*}

We also conducted exploratory logistic regression models within the ads posted to Facebook. We found that Facebook permitted 89\% park and holiday advertisements compared to 99\% of product ads, a statistically-significant difference of 10 percentage points (p=0.005) (Figure \ref{fig:issue_vs_candidate}, Table \ref{table:coefficients} ``Ad Type"). We also tested hypotheses about differences in leaning, ad location, and the location of the ad-poster. In each case, we did not observe a statistically-significant result, though it is possible that with a larger sample size, we may have done so (Table \ref{table:coefficients}).
% Based these exploratory findings, we created a combined of mistaken enforcement rates per platform based on election level and political leaning (Figure \ref{fig:state_leaning}).

\begin{table}
\begin{center}
%\resizebox{\linewidth}{!}{%
\begin{tabular}{l c c c c }
\hline
 & Ad Type & Leaning & Location & Ad Poster \\
\hline
(Intercept)              & $5.05^{***}$ & $2.77^{***}$ & $4.34^{***}$ & $4.06^{***}$ \\
                         & $(1.00)$     & $(0.39)$     & $(1.01)$     & $(0.71)$     \\
Ad Type (Park \& Parade) & $-2.97^{**}$ &              &              &              \\
                         & $(1.06)$     &              &              &              \\
Leaning (Republican)     &              & $0.88$       &              &              \\
                         &              & $(0.70)$     &              &              \\
Location (State)         &              &              & $-1.52$      &              \\
                         &              &              & $(1.06)$     &              \\
Poster Location (US)     &              &              &              & $-1.42$      \\
                         &              &              &              & $(0.80)$     \\
\hline
AIC                      & 72.62        & 85.25        & 83.99        & 83.06        \\
BIC                      & 79.56        & 92.20        & 90.93        & 90.00        \\
Log Likelihood           & -34.31       & -40.63       & -39.99       & -39.53       \\
Deviance                 & 68.62        & 81.25        & 79.99        & 79.06        \\
Num. obs.                & 238          & 238          & 238          & 238          \\
\hline
\multicolumn{5}{l}{\scriptsize{$^{***}p<0.001$, $^{**}p<0.01$, $^*p<0.05$}}
\end{tabular}%}
\caption{Logistic regression models testing univariate differences in publication rates based on ad type (Park \& Parade vs Product), Leaning (Republican vs Democrat) Location, (State vs Federal), and Ad Poster Location (US vs non-US)}
\label{table:coefficients}
\end{center}
\end{table}

% latex table generated in R 3.4.3 by xtable 1.8-2 package
% Sun Oct 14 00:01:44 2018
\begin{table}[h]
\centering
%\resizebox{\linewidth}{!}{%
\begin{tabular}{llllllrrrr}
\hline
% \textbf{platform} & \textbf{ad poster} & \textbf{location} & \textbf{leaning} & \textbf{ad type} & \textbf{\#} & \textbf{published} \\ 
%  \hline
% Facebook & US & federal & Democrat & candidate.mistake &  20 & 100.0\% \\ 
% Facebook & US & federal & Republican & candidate.mistake &  19 & 94.7\% \\ 
% Facebook & US & state & Democrat & candidate.mistake &  20 & 100.0\% \\ 
% Facebook & US & state & Democrat & issue.mistake &  20 & 75.0\% \\ 
% Facebook & US & state & Republican & candidate.mistake &  20 & 100.0\% \\ 
% Facebook & US & state & Republican & issue.mistake &  21 & 90.5\% \\ 
% Facebook & Non-US & federal & Democrat & candidate.mistake &  20 & 100.0\% \\ 
% Facebook & Non-US & federal & Republican & candidate.mistake &  19 & 100.0\% \\ 
% Facebook & Non-US & state & Democrat & candidate.mistake &  19 & 100.0\% \\ 
% Facebook & Non-US & state & Democrat & issue.mistake &  20 & 90.0\% \\ 
% Facebook & Non-US & state & Republican & candidate.mistake &  20 & 100.0\% \\ 
% Facebook & Non-US & state & Republican & issue.mistake &  20 & 100.0\% \\ 
% \hline
\textbf{platform} & \textbf{ad poster} & \textbf{location} & \textbf{leaning} & \textbf{ad type} & \textbf{\#} & \textbf{published} \\ 
 \hline
Facebook & US & federal & Democrat & candidate.mistake &  20 & 100.0\% \\ 
Facebook & US & federal & Republican & candidate.mistake &  19 & 94.7\% \\ 
Facebook & US & state & Democrat & candidate.mistake &  20 & 100.0\% \\ 
Facebook & US & state & Democrat & issue.mistake &  20 & 75.0\% \\ 
Facebook & US & state & Republican & candidate.mistake &  20 & 100.0\% \\ 
Facebook & US & state & Republican & issue.mistake &  21 & 90.5\% \\ 
Facebook & Non-US & federal & Democrat & candidate.mistake &  20 & 100.0\% \\ 
Facebook & Non-US & federal & Republican & candidate.mistake &  19 & 100.0\% \\ 
Facebook & Non-US & state & Democrat & candidate.mistake &  19 & 100.0\% \\ 
Facebook & Non-US & state & Democrat & issue.mistake &  20 & 90.0\% \\ 
Facebook & Non-US & state & Republican & candidate.mistake &  20 & 100.0\% \\ 
Facebook & Non-US & state & Republican & issue.mistake &  20 & 100.0\% \\ 
\hline
Google & US & federal & Democrat & candidate.mistake &  20 & 100.0\% \\ 
Google & US & federal & Republican & candidate.mistake &  19 & 100.0\% \\ 
Google & US & state & Democrat & candidate.mistake &  20 & 100.0\% \\ 
Google & US & state & Democrat & issue.mistake &  20 & 100.0\% \\ 
Google & US & state & Republican & candidate.mistake &  20 & 100.0\% \\ 
Google & US & state & Republican & issue.mistake &  21 & 100.0\% \\ 
Google & Non-US & federal & Democrat & candidate.mistake &  20 & 100.0\% \\ 
Google & Non-US & federal & Republican & candidate.mistake &  20 & 100.0\% \\ 
Google & Non-US & state & Democrat & candidate.mistake &  19 & 100.0\% \\ 
Google & Non-US & state & Democrat & issue.mistake &  20 & 100.0\% \\ 
Google & Non-US & state & Republican & candidate.mistake &  20 & 100.0\% \\ 
Google & Non-US & state & Republican & issue.mistake &  20 & 100.0\% \\ 
   \hline
\end{tabular}%}
\label{table:observed_ad_rates}
\caption{Number of ads placed and percentage of ads published, for each combination of platform, investigator location, election location, leaning, and type of advertisement.}
\end{table}

Taken together, our audit demonstrated that Facebook was systematically prohibiting non-election material of importance to American civic life, including public holidays and government websites. Further reports and journalism corroborated our findings, showing that prohibitions on government advertising affected housing and urban development departments \cite{smith_facebook_2019}. When scaled across an entire society these errors could have significant impacts on civic life and people's access to government services. While we cannot confirm the role of our research in corporate policy changes, Facebook altered its policies to accommodate government advertising in 2019 \cite{ortutay_facebook_2019}. As is common with social scientific audit studies \cite{quillian_meta-analysis_2017}, any evaluation of Facebook's changes to its decision-making system would require further audits, which could be repeated with our software and audit study design.

\subsection{Limitations}
All enforcement systems make mistakes. Our audit study shows the rates at which Facebook and Google prohibited non-election ads under their political advertising policies during the 2018 U.S. midterm elections.

Overall, our study discovers evidence of systematic mistakes (false positives) by Facebook, who prevented the publication of ads that were acceptable within the company's own policies. Facebook regularly blocked ads for national holidays, government national park websites, and non-partisan products that happen to share common characters with a candidate name. This false positive rate of 4.2\% is not representative of all advertisements posted to Facebook, but it does represent an important segment of online advertising in the United States. 

This audit study found no evidence of political ad policy enforcement by Google. This audit focuses on decisions that are too restrictive and, for legal reasons, does not study decisions that are too permissive. For that reason, we cannot provide guidance on whether our finding was due to a general lack of political policy enforcement by the company, whether the company has narrower internal policies, or whether Google was more accurate at enforcing its policies than Facebook in the cases we examined.

This study has several limitations. First, we have no information about either company's enforcement of ads that genuinely violate their policies. Second, our findings only generalize to the kinds of ads we tested: it is possible that companies made more mistakes with other kinds of ads that have received attention in the press, such as news articles and LGBTQ gatherings. Third, failures to find differences in publication rates should not be interpreted as proof of no difference; a larger study may have detected differences more clearly. Fourth, in this study, we offered 1 unit of currency (such as one US or Canadian dollar) per day per ad. If platforms offered greater scrutiny to ads that involve more money, it is possible that the rate of mistakes might be higher or lower in those cases. Finally, since platforms frequently change policies, internal guidelines, and training procedures with little public notice, these findings are most strongly informative about the period we studied.

We note that our audit study was not completely automated. Our software enabled us scrape images and text for certain ad types, generate ads based on templates, conduct a power analysis, and allocate ads to testers. Human testers were essential to this audit study, since we wanted to rule out platform policies about inauthentic activity. In other types of audits, researchers might be more able to use of platform-provided APIs to automatically place ads, enabling them to achieve greater dimensional complexity in their audit studies.

\section{Discussion}
We have described common design challenges that constrain the validity of audit studies, field experiments that contribute pragmatic and scientific knowledge about decision-making system errors. We have described areas where software could guide the design of these studies, broaden their dimensional complexity, and make them more efficient. To explore and validate these ideas, we prototyped software to support an audit of Google and Facebook's political ad policies during the 2018 U.S. midterm election. Using a study design informed by our software and using prompts auto-generated by that software, we were able to observe systematic errors in Facebook's human and machine decision-making about political ads. We also learned wider lessons for designing software to support audit studies.

% We created software to (a) support variable selection, (b) generate audit materials, (b) allocate audit attempts to testers, and (d) generate pre-registered statistical results and illustrations.

The trade-off between validity and dimensional complexity is a central design challenge for any audit study. In this project, we identified several areas where software can broaden the dimensional complexity of an audit and guide decisions about those trade-offs. By auto-generating prompts with data from APIs, researchers can reduce the labor of creating them manually while ensuring consistency. While our system to allocate prompts to testers was simple, we expect that software for recruiting and coordinating testers could also substantially expand the potential complexity of audit studies. We also found that software simulations can guide researcher decisions about sample sizes and the validity-complexity trade-off.

Despite these gains in complexity, some constraints cannot be addressed with software. The first is budget. Since our audit involved publishing ads, the financial cost of our audit study was a linear function of our sample size, determined by the statistical power we needed. Second, we limited our prompts to kinds of content that could be generated from publicly-available data. The potential for automating audit materials for a given system will be related to the availability of such data sources. Finally, for audits that involve active tester involvement, training and ongoing coordination are further limits on the complexity of any audit study.

Some of these constraints might be overcome through future work in automated audit studies. For example, software that acts on behalf of testers to carry out audit procedures may be able to reduce the complexities of training and coordinating testers-- without reducing the realism of the audit. Future audits could reduce sample sizes through sequential testing algorithms that incorporate information from each new observation into models that auto-allocate new testers to post new prompts. Even without sequential testing, audit study researchers could use targeted advertising and matching algorithms to recruit and allocate audit prompts across multi-dimensional samples of participants.

Audit studies contribute valuable knowledge to science and society. In democracies, the public knowledge from audit studies is valuable for holding power accountable and correcting errors. By supporting the audit process with software, researchers can broaden the validity and value of this valuable research method.

\section*{Conflicts of Interest}
We have no conflicts of interest to report with Facebook and Google.

\section*{Acknowledgments}
We are grateful to Molly Sauter, who provided logistical support to this study, and to Jon Penney, who provided helpful advice and feedback.
We are also grateful to Chris Peterson, Melissa Hopkins, Jason Griffey, and Ben Werdmuller for participating in our study as testers.
This research project was supported financially by the Princeton University Center for Information Technology Policy.

\bibliographystyle{ACM-Reference-Format}
\bibliography{main}

\end{document}